%% file: main.tex
\documentclass[conference]{IEEEtran}
\IEEEoverridecommandlockouts
\usepackage{latexsym}
\usepackage{graphicx}
\usepackage{amsfonts,amssymb,amsmath}
\def\BibTeX{{\rm B\kern-.05em{\sc i\kern-.025em b}\kern-.08em T\kern-.1667em\lower.7ex\hbox{E}\kern-.125emX}}

\usepackage{algorithmic}
\usepackage{algorithm}
\usepackage{subfig}
\usepackage{siunitx}
\usepackage{cite}

\usepackage[acronym]{glossaries}
\newcommand{\newac}{\newacronym}

\makeglossaries

\newac{los}{\text{LoS}}{line-of-sight}
\newac{nlos}{\text{NLoS}}{non-line-of-sight}

\input{math_lib}
\usepackage{anyfontsize}

\begin{document}

\title{UAV Trajectory Optimization and Tracking for User Localization in Wireless Networks
}

\author{
\IEEEauthorblockN{Omid Esrafilian, Rajeev Gangula, and David Gesbert}
\IEEEauthorblockA{Communication Systems Department, EURECOM, Sophia Antipolis, France\\
\text{\{esrafili, gangula, gesbert\}@eurecom.fr}
}
}

\maketitle

\begin{abstract} 

In this paper, we investigate the problem of UAV-aided user localization in wireless networks. Unlike the existing works, we do not assume perfect knowledge of the UAV location, hence we not only need to localize the users but also to track the UAV location. To do so, we utilize the time-of-arrival along with received signal strength radio measurements collected from users using a UAV. A simultaneous localization and mapping (SLAM) framework building on the Expectation-Maximization-based least-squares method is proposed to classify measurements into line-of-sight or non-line-of-sight categories and learn the radio channel, and at the same, localize the users and track the UAV. This framework also allows us to exploit other types of measurements such as the rough estimate of the UAV location available from GPS, and the UAV velocity measured by an inertial measurement unit (IMU) on-board, to achieve better localization accuracy. Moreover, the trajectory of the UAV is optimized which brings considerable improvement to the localization performance. The simulations show the out-performance of the developed algorithm when compared to other approaches.

\end{abstract}


\input{sections/01intro}

\input{sections/02system}
\input{sections/03LocalizationLearning}
\input{sections/04TrajectoryOptimization}

\input{sections/05results}

\section{Conclusion}\label{sec:conclusion}

This paper considered the problem of simultaneous UAV tracking as a mobile anchor and localization of ground users. To do so, we utilize the ToA along with RSS radio measurements collected from users using a UAV. An EM-based SLAM framework was proposed to classify measurements into LoS/NLoS categories and learn the radio channel, and simultaneously, localize the users and track the UAV. Using this framework we are able to exploit other types of measurements such as the estimate of the UAV location available from GPS, and the UAV velocity measured by an IMU onboard the UAV for better localization. The trajectory of the UAV was also optimized to collect the most informative measurements which results in more accurate localization.

\section{Acknowledgments}
This work was funded via the HUAWEI
France supported Chair on Future Wireless Networks at EURECOM.

\input{sections/06Appendix}

\bibliographystyle{IEEEtran}
\bibliography{literature.bib}

\end{document}

%% file: math_lib.tex
\DeclareMathAlphabet{\mathbit}{OML}{cmr}{bx}{it}
\DeclareMathAlphabet{\mathsf}{OT1}{cmss}{m}{n}
\DeclareMathAlphabet{\mathTXf}{OT1}{cmss}{bx}{it}

\DeclareMathOperator{\trace}{tr}

\DeclareMathOperator{\Transpose}{T}

\DeclareMathOperator{\Exp}{E}

\DeclareMathAlphabet{\mathpzc}{OT1}{pzc}{m}{it}

\newcommand{\Tr}{{\Transpose}}



%% file: sections/01intro.tex
\section{Introduction}\label{sec:Intro}

In a wireless localization system, nodes with perfectly-known positions known as anchor nodes (which can be stationary or mobile) collect various radio measurements from the emitted radio frequency (RF) signals from the users in the network and use them for localization purposes. Various measurements  such  as  received signal strength (RSS), time-of-arrival (ToA), angle of arrival (AoA), time-difference-of-arrival (TDoA), etc.,  can  be  obtained from the RF signals by the anchor nodes \cite{zekavat2011handbook,delRauLop}.

On the other hand, advancements in robotic technologies and miniaturization of wireless equipment have made it possible to have flying radio networks (FRANs), where wireless connectivity to ground users can be provided by aerial base stations (BSs) or relays that are mounted on unmanned aerial vehicles  (UAVs) \cite{GanEsaGes,MozSaadBennNamDebb}. The advantage of FRANs includes fast and dynamic network deployment during an emergency or temporary crowded events, providing connectivity in areas lacking network infrastructure, etc. While in terrestrial radio access networks static BSs are used as anchor nodes, in FRANs UAV BSs can be used as mobile anchor nodes \cite{bisio2021localization, albanese2021first}.
However, when it comes to aerial anchors, the location of the UAV is of crucial importance. Unfortunately, the UAV location is not precisely known and is subject to noise. Therefore, when using aerial mobile anchors, the problem becomes not only localizing the users but also tracking the UAV location \cite{yang2021survey}.

Localization of ground users using radio measurements collected by 
aerial UAV anchor nodes has recently gained  interest \cite{ zhou4027231flying, annepu2020unmanned, esrafilian20203d, esrafilian2022uav, liang2022uav, le2020hybrid, sinha2022impact, SinYapISm}.
The main advantage of using UAV anchors in localization compared to static anchors is that UAVs with their inherent 3D mobility can collect radio measurements in different geographic locations which improves the localization performance. In other words, the UAV in a different location can be considered a virtual static anchor.

The problem of ground users localization exploiting RSS measurements collected by 
aerial UAV anchor nodes
has been studied in \cite{zhou4027231flying, annepu2020unmanned, esrafilian20203d, esrafilian2022uav}. One common assumption in all these works is that the perfect knowledge of the UAV location is available. In \cite{esrafilian20203d}, the UAV jointly learns the radio channel and localizes the ground users. The trajectory of the UAV is also optimized to improve the localization accuracy. In \cite{zhou4027231flying}, the trajectory of the UAV is optimized to localize the users. A maximum likelihood estimation method is used to calculate the coordinate of unknown radio nodes. In \cite{annepu2020unmanned, esrafilian2022uav}, a neural network is used to estimate the ground users' location by taking into account the uncertainty which cannot be captured using classical approaches. 

The authors in \cite{liang2022uav} considered multi-UAV-aided localization systems where a combination of ToA and AoA measurements are used to localize ground users. The deployment of UAVs is also optimized for further improvement of localization performance. In \cite{le2020hybrid}, a hybrid ToA along with 1D AoA localization approach that merely requires
elevation AoA estimations to combine with ToA measurements is proposed. The impact of the antenna radiation pattern for the channel between the UAV as an aerial anchor and the ground users in a 3D localization system using time-based (ToA and TDoA) measurements has been studied in \cite {sinha2022impact, SinYapISm}.

In this paper, we propose a new algorithm for tracking the location of the UAV and at the same time localizing the users. To solve this problem, we exploit the ToA along with RSS measurements. We proposed an Expectation-Maximization -based (EM) localization algorithm employing a least-squares simultaneous localization and mapping (SLAM) framework to jointly classify measurements into line-of-sight (LoS) or non-line-of-sight (NLoS) categories and learn the radio channel, and at the same, localize the users and track the UAV. Additional measurements such as the UAV location estimate available from the global positioning system (GPS), and the UAV velocity measured by an inertial measurement unit (IMU) onboard is used to improve the localization accuracy.

To the best of our knowledge, simultaneous localization of users and tracking a mobile flying anchor by exploiting ToA and RSS measurements has not been studied in the literature. Specifically, our contributions are as follows:
\begin{itemize}
\item  A mobile-anchor-based localization algorithm is proposed by exploiting the ToA and RSS measurements.

\item A multi-modal least-squares-based SLAM problem is formulated for jointly classifying radio measurements, localizing the users, and at the same time tracking the mobile anchor.  

\item The trajectory of the mobile anchor is optimized to further improve the localization accuracy
\end{itemize}

%% file: sections/02system.tex
\begin{figure}[t]
\begin{centering}
\includegraphics[width=0.7\columnwidth]{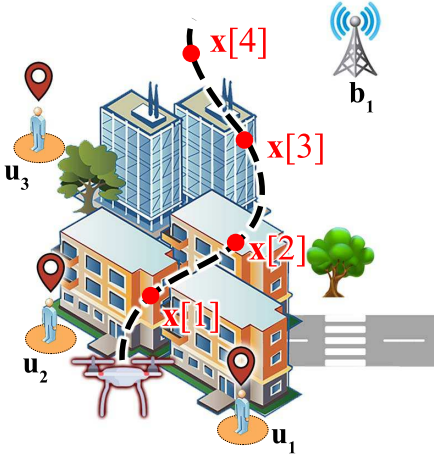}
\par\end{centering}
\caption{UAV-aided ground user localization system.\label{fig:SystemModel}}
\end{figure}

\section{System Model and Problem Formulation}\label{sec:SysModel}
We consider a scenario similar to the one illustrated in Fig. \ref{fig:SystemModel}, where $M$ static terrestrial base stations (BSs) and a UAV-relay that are capable of measuring radio signals from $K$ ground level users in a given service area. 
The terrestrial BSs are connected to both the UAV-relay and 
ground users. We assume that the location of the static BSs are known. The location of the $m$-th BSs are denoted by ${\bf{b}}_m=[x_m,y_m,z_m]^{\Tr}\in\mathbb{R}^{3}, m\in [1, M]$.
The users are
spread over the area and ${\bf u}_{k}=[x_{k},y_{k}]^{\Tr}\in\mathbb{R}^{2},\,k\in[1,K]$
denotes the $k$-th user's location. The users are considered static and their locations are unknown. 
The aim of this work is to estimate the user locations based on the radio measurements collected by the UAV-relay and the BSs.

\subsection{UAV Model}
We assume that the UAV mission lasts for a 
duration of $T$ seconds. We discretize the UAV mission time into $N$ equal time steps such that the UAV velocity is assumed to be constant within each time step. In the $n$-th time step, the UAV-relay/drone position is denoted by 
${\bf{x}}[n]=[x[n],y[n],z[n]]^{\Tr}\in\mathbb{R}^{3},\,n\in[1,N]$.
We assume that the UAV is equipped with a GPS receiver, the UAV location obtained using the GPS is given by

\begin{equation}
    {\Hat{\bf{x}}}[n] = {\bf{x}}[n] + {\bf{e}}_x,
\end{equation}
where ${\bf{e}}_x$ is the GPS measurement noise which is modeled by a Gaussian random variable with $\mathcal{N}(0,\sigma_{gps}^{2} * {\bf{I}}_2)$, where ${\bf{I}}_{2}$ is the identity matrix of size $2 \times 2$. The UAV velocity is also available from an IMU onboard the UAV and is denoted by 
\begin{equation}
    {\Hat{\bf{v}}}[n] = {\bf{v}}[n] + {\bf{e}}_v, \label{eq:uav_IMU_velocity}
\end{equation}
where ${\bf{v}}[n]$ is the true UAV velocity at time step $n$, and ${\bf{e}}_v$ is the velocity measurement noise which is modeled by a Gaussian random variable as $\mathcal{N}(0,\sigma_{vel}^{2} * {\bf{I}}_2)$. Assuming that the UAV moves with constant speed within each time step, the UAV velocity can be computed as follows 
\begin{equation}
    {\bf{v}}[n] = \frac{{\bf{x}}[n] - {\bf{x}}[n-1]}{\Delta t},
\end{equation}
where $\Delta t = \frac{T}{N}$. Therefore, \eqref{eq:uav_IMU_velocity} can be reformulated as

\begin{equation}
    {\Hat{\bf{v}}}[n] = \frac{{\bf{x}}[n] - {\bf{x}}[n-1]}{\Delta t} + {\bf{e}}_v. \label{eq:uav_IMU_velocity_2}
\end{equation}

\subsection{Channel Model}
We now describe the radio channel model between the BS, UAV and ground users. BSs and the UAV-relay have access to two types of radio measurements, namely, RSS (or channel gain) and ToA. 

Classically, the channel gain (in dB) between UAV position ${\bf{x}}[n]$ and user location ${\bf{u}}_k$ can be modeled as

\begin{equation}\label{eq:CH_gain_Model_dB}
\hat{g}_{k}[n] {=}
\begin{cases} 
  g_{k, \text{LoS}}[n] +\eta_{\text{LoS}}      & \small{\text{if} \text{ LoS}}\\
    g_{k, \text{NLoS}}[n] +\eta_{\text{NLoS}}       & \small{\text{if} \text{ NLoS}},
   \end{cases}
\end{equation} 

\noindent where $g_{k, s}[n] \triangleq  \beta_{s} + \alpha_{s}\, \phi_k[n]$, where
\begin{equation}\label{eq:log_dist_def}
    \phi_k[n]=\log_{10} \| {\bf{x}}[n] - {\bf{u}}_k\|,
\end{equation}
and $\alpha_{s}$ is the path loss exponent, $\beta_{s}$ is the
average channel gain at the reference point $d=1$ meter, $\eta_{s}$
denotes the shadowing component which is modeled as a Gaussian
random variable with $\mathcal{N}(0,\sigma_{s}^{2})$, and finally $s \in \mathcal{S} \triangleq \left\{ \text{LoS},\text{NLoS}\right\} $
emphasizes the strong dependence of the propagation parameters on
LoS or NLoS scenarios. Note that (\ref{eq:CH_gain_Model_dB}) represents
the channel gain which is averaged over the small scale fading of
unit variance. We assume that the channel parameters in \eqref{eq:CH_gain_Model_dB} are not known and need to be estimated from the measurements.
The probability distribution of a single measurement in \eqref{eq:CH_gain_Model_dB} is given by
\begin{equation} \label{eq:PDF_RSSI}
    p(g_{k}[n]) = (f_{k,\text{LoS}}[n])^{w_{k}[n]} (f_{k,\text{NLoS}}[n])^{(1-w_{k}[n])},
\end{equation}
where $\omega_{k}[n] \in \{0,1\}$ is the classifier binary variable (unknown) indicating whether a measurement falls into the LoS or NLoS category, and $f_{k,s}[n]$ has a Gaussian distribution with $\mathcal{N}(\beta_{s} + \alpha_{s}\, \phi_k[n] ,\sigma_{s}^{2})$.

The ToA measurement between the UAV at time step $n$ and the $k$-th user is modeled as follows

\begin{equation}\label{eq:CH_TOA_Model}
\hat{\tau}_{k}[n] {=}
\begin{cases} 
   \tau_{k}[n] + \xi_{\text{LoS}}      & \small{\text{if} \text{ LoS}}\\
    \tau_{k}[n] + \xi_{\text{NLoS}}     & \small{\text{if} \text{ NLoS}},
   \end{cases}
\end{equation} 

\noindent where 
\begin{equation*}
    \tau_{k}[n] = \frac{\| {\bf{x}}[n] - {\bf{u}}_k \|}{c},
\end{equation*}

\noindent where $c$ is the speed of light, and $\xi_s$ is the timing measurement noise which can be modeled as a Gaussian
random variable with $\mathcal{N}(\mu_{\tau,s},\sigma_{\tau, s}^{2})$. Note that the measurement noise is different for each segments of LoS/NLoS. $\mu_{\tau, s}$ reflects the error when there is no LoS link between the receiver and the transmitter, and is considered zero when $s = \text{LoS}$. 
The $\sigma_{\tau, s}^{2}$ similarly depends on the link status and also on the Bandwidth of the channel. Generally speaking, $\sigma_{\tau, s}^{2} \propto \frac{1}{\text{Bandwidth}} \frac{1}{\text{SNR}_{k, s}[n]}$. Where $\text{SNR}_{k, s}[n]$ is the signal-to-noise-ratio (SNR) of the signal between the UAV at time step $n$ and user $k$ at segment $s$. For simplicity, in this paper we assume an average SNR at each segment to model the ToA as follows
\begin{equation}
    \sigma_{\tau, s}^{2} \propto \frac{1}{\text{Bandwidth}} \frac{1}{\overline{\text{SNR}}_{s}},
\end{equation}
where $\overline{\text{SNR}}_{s}$ is the average SNR at each segment of LoS/NLoS. The probability distribution of a measurement in \eqref{eq:CH_TOA_Model} is given by
\begin{equation} \label{eq:PDF_ToA}
    p(\tau_{k}[n]) = (f_{\tau, k,\text{LoS}}[n])^{w_{k}[n]} (f_{\tau, k,\text{NLoS}}[n])^{(1-w_{k}[n])},
\end{equation}
where $f_{\tau, k,s}[n]$ has a Gaussian distribution with $\mathcal{N}(\frac{\| {\bf{x}}[n] - {\bf{u}}_k \|}{c} + \mu_{\tau,s} ,\sigma_{\tau, s}^{2})$. It is worth mentioning that, the same classification variable $w_{k}[n]$, as  used to classify measurements in \eqref{eq:PDF_RSSI}, is utilized to classify ToA measurements in \eqref{eq:PDF_ToA}. This stems from the fact that both measurements are taken at the same time and the status of the link is the same for both types of measurements.

The channel gain and ToA measurements between user $k$ and BS $m$ represented by $\hat{g}_{m, k}, \hat{\tau}_{m, k}$, and the channel gain and ToA measurements between the UAV at time step $n$ and BS $m$ are denoted by $\hat{g}_{m}[n], \hat{\tau}_{m}[n]$, respectively. 
These above observations follow similar link model akin to \eqref{eq:CH_gain_Model_dB} and \eqref{eq:CH_TOA_Model}. 
The probability distribution of measurements collected by BSs from the UAV and the users follow similar model in \eqref{eq:PDF_RSSI} and \eqref{eq:PDF_ToA}.

%% file: sections/03LocalizationLearning.tex
\section{User Localization and UAV Tracking\label{sec:LearningLocalization}}

In this section, we propose an algorithm to estimate the user locations from the radio measurements collected by the UAV and BSs. Let us denote an arbitrary set of measurements available at UAV (taken by the UAV and also by the BSs) during the mission as
$\mathcal{G} = \left\{o[n], \gamma_k[n], \gamma_m[n], \gamma_{m, k}; \forall n, m, k\right\}$, where $o[n]$ is the UAV odometry measurements at time step $n$, $\gamma_k[n]$ is a tuple of measurements collected by the UAV  at time step $n$ from user $k$, $\gamma_m[n]$ is a tuple of measurements collected by BS $m$ from the UAV at time step $n$, and $\gamma_{m, k}$ is a tuple of measurements collected by BS $m$ from user $k$ defined as follows
\begin{align*}
    o[n] &\triangleq ({\Hat{\bf{x}}}[n], {\Hat{\bf{v}}}[n]), n\in [1,N], \\
     \gamma_k[n] &\triangleq \left(\hat{g}_{k}[n], \hat{\tau}_{k}[n] \right), n\in [1,N],  k\in [1,K]\\
     \gamma_m[n] &\triangleq \left(\hat{g}_{m}[n], \hat{\tau}_{m}[n] \right), n\in [1,N], m\in [1,M],\\
     \gamma_{m, k} &\triangleq \left(\hat{g}_{m, k}, \hat{\tau}_{m, k} \right), m\in [1,M], k\in [1,K].
\end{align*}
\noindent Note that the LoS/NLoS status of channel gain and ToA measurements (measurements between the UAV and users, between the UAV and the BSs, and between users and the BSs) are not available and need to be estimated as well. It is also worth mentioning that, the true location and the velocity of the UAV are not available (i.e. the GPS and the IMU measurements are subject to the noise), therefore we not only have to localize the users but also need to track the UAV location.

Assuming that collected measurements conditioned on the channel and user positions are independent and identically distributed (i.i.d), the negative log-likelihood of measurements can be written as
\begin{equation}\label{eq:all_log_lkelihood}
    \mathcal{L} = \mathcal{L}_{\text{UAV}} + \mathcal{L}_{\text{UAV-UE}} + \mathcal{L}_{\text{BS-UAV}} + \mathcal{L}_{\text{BS-UE}},
\end{equation}
\noindent where $\mathcal{L}_{\text{UAV}}$ is the loss function regarding the odometry measurements, $\mathcal{L}_{\text{UAV-UE}}$ is the loss function for measurements taken by the UAV from users, $\mathcal{L}_{\text{BS-UAV}}$ is the loss function for measurements taken by BSs from the UAV, and  $\mathcal{L}_{\text{BS-UE}}$ is the loss function for the measurements taken by BSs from the users. The loss functions are defined as follows

{\small
\begin{equation*}
\begin{aligned}
    \mathcal{L}_{\text{UAV}} \triangleq& \sum_{n=1}^{N}
  \frac{1}{\sigma_{gps}^2}\left \| {\Hat{\bf{x}}}[n] - {\bf{x}}[n] \right\|^2 + \\
  & \sum_{n=2}^{N}
  \frac{1}{\sigma_{vel}^2}\left \| {\Hat{\bf{v}}}[n] - \frac{{\bf{x}}[n] - {\bf{x}}[n-1]}{\Delta t} \right\|^2, 
  \end{aligned}
\end{equation*}

\begin{equation*}
\begin{aligned}
    \mathcal{L}_{\text{UAV-UE}} \triangleq& \log\left(\frac{\sigma_{\text{LoS}}^2}{\sigma_{\text{NLoS}}^2} \frac{\sigma_{\tau, \text{LoS}}^2}{\sigma_{\tau,\text{NLoS}}^2} \right)\sum_{n=1}^{N} \sum_{k=1}^{K} \omega_{k}[n] + \\
    &\sum_{n=1}^{N} \sum_{k=1}^{K} 
  \frac{\omega_{k}[n]}{\sigma_{\text{LoS}}^2}\left |\hat{g}_{k}[n] - \beta_{\text{LoS}}-\alpha_{\text{LoS}}\, \phi_k[n]\right|^2 +\\
  &\sum_{n=1}^{N} \sum_{k=1}^{K} 
  \frac{1 - \omega_{k}[n]}{\sigma_{\text{NLoS}}^2}\left |\hat{g}_{k}[n] - \beta_{\text{NLoS}}-\alpha_{\text{NLoS}}\, \phi_k[n]\right|^2 +\\
 &\sum_{n=1}^{N}\sum_{k=1}^{K}
  \frac{\omega_{k}[n]}{\sigma_{\tau, \text{LoS}}^2}\left | \hat{\tau}_{k}[n] - \frac{\| {\bf{x}}[n] - {\bf{u}}_k \|}{c} - \mu_{\tau, \text{LoS}}\right|^2 + \\
   &\sum_{n=1}^{N}\sum_{k=1}^{K}
  \frac{1 - \omega_{k}[n]}{\sigma_{\tau, \text{NoS}}^2}\left | \hat{\tau}_{k}[n] - \frac{\| {\bf{x}}[n] - {\bf{u}}_k \|}{c} - \mu_{\tau, \text{NLoS}}\right|^2,
  \end{aligned}
\end{equation*}

\begin{equation*}
\begin{aligned}
    \mathcal{L}_{\text{BS-UAV}} \triangleq &\log\left(\frac{\sigma_{\text{LoS}}^2}{\sigma_{\text{NLoS}}^2} \frac{\sigma_{\tau, \text{LoS}}^2}{\sigma_{\tau,\text{NLoS}}^2} \right)\sum_{n=1}^{N} \sum_{m=1}^{M}\omega_{m}[n] + \\
    &\sum_{n=1}^{N} \sum_{m=1}^{m} 
  \frac{\omega_{m}[n]}{\sigma_{\text{LoS}}^2}\left |\hat{g}_{m}[n] - \beta_{\text{LoS}}-\alpha_{\text{LoS}}\, \phi_m[n]\right|^2 +\\
  &\sum_{n=1}^{N} \sum_{m=1}^{M} 
  \frac{1 - \omega_{m}[n]}{\sigma_{\text{NLoS}}^2}\left |\hat{g}_{m}[n] - \beta_{\text{NLoS}}-\alpha_{\text{NLoS}}\, \phi_m[n]\right|^2 +\\
 &\sum_{n=1}^{N}\sum_{m=1}^{M}
  \frac{\omega_{m}[n]}{\sigma_{\tau, \text{LoS}}^2}\left | \hat{\tau}_{m}[n] - \frac{\|{\bf{b}}_m - {\bf{x}}[n] \|}{c} - \mu_{\tau, \text{LoS}}\right|^2 + \\
   &\sum_{n=1}^{N}\sum_{m=1}^{M}
  \frac{1 - \omega_{m}[n]}{\sigma_{\tau, \text{NoS}}^2}\left | \hat{\tau}_{m}[n] - \frac{\|{\bf{b}}_m - {\bf{x}}[n] \|}{c} - \mu_{\tau, \text{NLoS}}\right|^2,
  \end{aligned}
\end{equation*}

\begin{equation*}
\begin{aligned}
    \mathcal{L}_{\text{BS-UE}} \triangleq &\log\left(\frac{\sigma_{\text{LoS}}^2}{\sigma_{\text{NLoS}}^2} \frac{\sigma_{\tau, \text{LoS}}^2}{\sigma_{\tau,\text{NLoS}}^2} \right)\sum_{m=1}^{M} \sum_{k=1}^{K}\omega_{m, k} + \\
    &\sum_{m=1}^{M} \sum_{k=1}^{K} 
  \frac{\omega_{m, k}}{\sigma_{\text{LoS}}^2}\left |\hat{g}_{m, k} - \beta_{\text{LoS}}-\alpha_{\text{LoS}}\, \phi_{m, k}\right|^2 +\\
  &\sum_{m=1}^{M} \sum_{k=1}^{K} 
  \frac{1 - \omega_{m, k}}{\sigma_{\text{NLoS}}^2}\left |\hat{g}_{m, k} - \beta_{\text{NLoS}}-\alpha_{\text{NLoS}}\, \phi_{m, k}\right|^2 +\\
 &\sum_{m=1}^{M} \sum_{k=1}^{K}
  \frac{\omega_{m, k}}{\sigma_{\tau, \text{LoS}}^2}\left | \hat{\tau}_{m, k} - \frac{\|{\bf{b}}_m - {\bf{u}}_k \|}{c} - \mu_{\tau, \text{LoS}}\right|^2 + \\
   &\sum_{m=1}^{M} \sum_{k=1}^{K}
  \frac{1 - \omega_{m, k}}{\sigma_{\tau, \text{NoS}}^2}\left | \hat{\tau}_{m, k} - \frac{\|{\bf{b}}_m - {\bf{u}}_k \|}{c} - \mu_{\tau, \text{NLoS}}\right|^2, 
  \end{aligned}
\end{equation*}
}

\noindent where $\omega_{m}[n], \omega_{m, k}$ are the binary classification variables determining the status of measurements taken by the BSs from the UAV, and the users, respectively. The terms  $\phi_m[n], \phi_{m, k}$ are defined similar to \eqref{eq:log_dist_def}. 

The estimate of the unknown user and the UAV locations can then be obtained by solving
\begin{equation}\label{eq:SLAM_Opt_Org}
           \min_{\substack{{\bf{x}}[n],\,{{\bf{u}}}_k, \mathcal{\theta}, \mathcal{W} \\ \forall n, k}} \quad \mathcal{L},
\end{equation}
\noindent where $\mathcal{\theta} = \left \{ \alpha_{\text{s}}, \beta_{\text{s}}, \sigma_{s}^2, \sigma_{\tau, s}^2, \mu_{\tau, s}; s \in \{\text{LoS}, \text{NLoS}\} \right\}$, and $\mathcal{W} = \{ \omega_{k}[n], \omega_{m}[n], \omega_{m, k}; \forall n, m, k\}$. Solving problem \eqref{eq:SLAM_Opt_Org} is challenging, since it is a mixed-integer problem of simultaneous user localization, UAV tracking, channel learning, and measurements classification. In the following sections, we introduce an iterative algorithm to jointly classify the measurements and learn the channel parameters, and at the same time localize the users and track the UAV. At each iteration of the algorithm we employ an EM algorithm to classify the measurements and learn the channel, and then we use a least-square-based SLAM framework to localize the users and track the UAV.

\subsection{Learning and Classification}
At this phase of the algorithm, we fix the user and the UAV location and we solve the problem \eqref{eq:SLAM_Opt_Org} to find the classification variables and also to learn the channel. Moreover, since the classification variables are the same for both channel gain and the ToA measurements, without loss of optimally, we classify the measurements only considering the channel gains. For the sake of clarity, let's only consider the channel gain measurements between the UAV and the users. Therefore, the log likelihood of the channel gain measurements can be rewritten as follows

 \begin{equation}\label{eq:EM_RSS_log_likelihood}
 \begin{aligned}
      \mathcal{L}_{\text{UAV-UE}}^{g} =& \sum_{n, k} \log p\left( \hat{g}_{k}[n]; \theta^{g}\right) \\
      = &\sum_{n, k} \log \sum_{w_{k}[n]\in \mathcal{S}} p\left( \hat{g}_{k}[n], w_{k}[n]; \theta^{g}\right),
 \end{aligned}
 \end{equation}
 
 \noindent where $\mathcal{L}_{\text{UAV-UE}}^{g}$ indicates the log-likelihood for only channel gain measurements between the UAV and the users, and $\theta^{g} = \left \{ \alpha_{\text{s}}, \beta_{\text{s}}, \sigma_{s}^2; s \in \{\text{LoS}, \text{NLoS}\} \right\}$. For each measurement, let $q(w_{k}[n])$ be a distribution over $w_{k}[n]$. We can then reformulate \eqref{eq:EM_RSS_log_likelihood} as

\begin{equation}\label{eq:EM_RSS_jensen}
 \begin{aligned}
      \mathcal{L}_{\text{UAV-UE}} = &\sum_{n, k} \log \sum_{w_{k}[n]\in \mathcal{S}} q(w_{k}[n]) \frac{p( \hat{g}_{k}[n], w_{k}[n];\theta^{g})}{q(w_{k}[n])}\\
      \overset{(a)}{\ge } & \sum_{n, k} \sum_{w_{k}[n]\in \mathcal{S}} q(w_{k}[n]) \log \frac{p( \hat{g}_{k}[n], w_{k}[n];\theta^{g})}{q(w_{k}[n])},
 \end{aligned}
 \end{equation}
 where step $(a)$ follows from Jensen's inequality which is a lower bound on the original likelihood. The lower bound can be tighten by choosing $q$ as follows
 
\begin{equation}\label{eq:E_step}
\begin{aligned}
    q(w_{k}[n]=s) =& \frac{p(\hat{g}_{k}[n]| w_{k}[n]=s;\theta^{g}) p(w_{k}[n]=s)}{\sum_{s'\in \mathcal{S}} p(\hat{g}_{k}[n]| w_{k}[n]=s';\theta^{g}) p(w_{k}[n]=s')}\\
    = & \frac{p(\hat{g}_{k}[n]| \,s;\theta^{g}) \pi_{s}}{\sum_{s'\in \mathcal{S}} p(\hat{g}_{k}[n]| \, s';\theta^{g}) \pi_{s'}}, 
      \end{aligned}
\end{equation}

Let's denote $\Omega_{k, s}[n] = q(w_{k}[n]=s)$, therefor the maximum log-likelihood estimation of parameters $\theta^{g}$ can be obtained by solving the following problem
\begin{subequations}\label{eq:MLE_Problem}
\begin{align}
  \begin{split}
    \max_{\theta^{g}, \pi_{s}} & \sum_{n, k} \sum_{s\in \mathcal{S}} \Omega_{k, s}[n] \log \frac{p( \hat{g}_{k}[n]| \, s;\theta^{g}) \pi_{s}}{\Omega_{k, s}[n]}\end{split}\\
    \begin{split} \label{eq:sum_to_one}
    \text{s.t.}& \: \sum_{s\in \mathcal{S}} \pi_{s} = 1.
\end{split}
      \end{align}
\end{subequations}

This problem is non-convex, therefore, we find a sub-optimal solution by solving \eqref{eq:MLE_Problem} iteratively. This algorithm iterates between two steps known as expectation and maximization (E-M steps). During the E-step, the $\Omega_{k, s}[n]$ is computed according to \eqref{eq:E_step} while fixing parameters $\theta^{g}$. In the M-step, problem \eqref{eq:MLE_Problem} is solved only for parameters $\theta^{g}$ by fixing $\Omega_{k, s}[n]$. We denote $i$ as the iteration index of the E-M algorithm and the process is repeated for $I$ iterations.

Let $\theta^{g, (i-1)}, \pi_{s'}^{(i-1)}$ be the parameters available from the $(i-1)$-th iteration, then during the E-step we have
\begin{equation}
    \Omega_{k, s}^{(i)}[n] = \frac{p(\hat{g}_{k}[n]| \,s;\theta^{g, (i-1)}) \pi_{s}^{(i-1)}}{\sum_{s'\in \mathcal{S}} p(\hat{g}_{k}[n]|  \,s';\theta^{g, (i-1)}) \pi_{s'}^{(i-1)}}.
\end{equation}
For the M-step, \eqref{eq:MLE_Problem} can be reformulated as follows:
\begin{subequations} \label{eq:EM_M_step}
\begin{align}
  \begin{split} \label{eq:EM_M_step_Obj}
    \max_{\theta^{g, (i)}, \pi_{s}^{(i)}} & \sum_{n, k} \sum_{s\in \mathcal{S}} \Omega_{k, s}^{(i)}[n] \log \frac{p( \hat{g}_{k}[n]|  \,s;\theta^{g, (i)}) \pi_{s}^{(i)}}{\Omega_{k, s}^{(i)}[n]}\end{split}\\
    \begin{split}
    \text{s.t.}& \: \eqref{eq:sum_to_one}.
\end{split}
      \end{align}
\end{subequations}
The objective function \eqref{eq:EM_M_step_Obj} is concave. By setting the derivative of \eqref{eq:EM_M_step} with respect to ${\theta^{g, (i)}, \pi_{s}^{(i)}}$ to zero and solving, we find

\begin{equation}
    \begin{bmatrix}
    \alpha_s^{(i)}\\
    \beta_s^{(i)}
    \end{bmatrix} = {\bf{A}}_{s}^{-1}\,{\bf{b}}_{s},
\end{equation}
where
\begin{equation*}
    {\bf{A}}_{s} = \begin{bmatrix}
  \sum_{n,k} \Omega_{k, s}^{(i)}[n]\, \phi_k^2[n] & \sum_{n,k} \Omega_{k, s}^{(i)}[n]\, \phi_k[n]\\
  \sum_{n,k} \Omega_{k, s}^{(i)}[n]\, \phi_k[n] & \sum_{n,k} \Omega_{k, s}^{(i)}[n]
  \end{bmatrix},
\end{equation*}
\begin{equation*}
    {\bf{b}}_{s} = \begin{bmatrix}
  \sum_{n,k} \Omega_{k, s}^{(i)}[n]\, \phi_k[n]\,\hat{g}_{k}[n]\\
  \sum_{n,k} \Omega_{k, s}^{(i)}[n]\,\hat{g}_{k}[n]
  \end{bmatrix},
\end{equation*}
\noindent and 
\begin{equation}
    {\sigma}_{s}^{(i)} = \sqrt{\frac{\sum_{n,k}  \Omega_{k, s}^{(i)}[n] \left (\hat{g}_{k}[n] - \beta_{s}^{(i)}-\alpha_{s}^{(i)}\, \phi_k[n]\right )^2}{N\, K}}.
\end{equation}
\noindent The value of $\pi_s^{(i)}$ is given by
\begin{equation}
    \pi_s^{(i)} = \frac{\sum_{n,k}\Omega_{k, s}^{(i)}[n]}{N \, K}.
\end{equation}

Having classified the measurements, the $\sigma_{\tau, s}^2, \mu_{\tau, s}; s \in \{\text{LoS}, \text{NLoS}\}$ can be found as follows

\begin{equation} \label{eq:ToA_opt}
           \min_{\substack{\sigma_{\tau, s}^2, \mu_{\tau, s} \\ s \in \mathcal{S}}} \quad \mathcal{L}_{\text{UAV-UE}}^{\tau},
\end{equation}
\noindent where $\mathcal{L}_{\text{UAV-UE}}^{\tau}$ indicates the log-likelihood for only ToA measurements between the UAV and the users. Solving \eqref{eq:ToA_opt} for $\sigma_{\tau, s}^2, \mu_{\tau, s}$ we can obtain

\begin{equation}
    \mu_{\tau, s} = \frac{\sum_{n,k}\Omega_{k, s}^{(I)}[n] (\hat{\tau}_{k}[n] - \frac{\| {\bf{x}}[n] - {\bf{u}}_k \|}{c}) }{N \, K},
\end{equation}
and 
\begin{equation}
    {\sigma}_{\tau, s} = \sqrt{\frac{\sum_{n,k}  \Omega_{k, s}^{(I)}[n] \left (\hat{\tau}_{k}[n] - \frac{\| {\bf{x}}[n] - {\bf{u}}_k \|}{c} - \mu_{\tau, \text{LoS}}\right )^2}{N\, K}},
\end{equation}
\noindent where the superscript $(I)$ indicates the value available from the final iteration of EM algorithm.
Finally, to obtain the labels for each measurement, we user the hard classification as follows
\begin{equation}
w_{k}[n] =
\begin{cases} 
      1     & \quad \text{if}\,\, \Omega_{k, s}^{(I)}>0.5\\
         0     & \quad \text{else}
   \end{cases}.
\end{equation}

Similar to above analysis, the E-M algorithm can be solved for $\mathcal{L}$ defined in  \eqref{eq:all_log_lkelihood} by considering all the collected measurements to estimate all classification variables $\mathcal{W}$, for the sake of the limited space we skip the details.

\subsection{User Localization and UAV Tracking} \label{sec:LS_SLAM}
Once we know the classification variables $\mathcal{W}$, we continue to find the location of the users and track the UAV. The optimization \eqref{eq:SLAM_Opt_Org} can now rewritten as

\begin{equation}\label{eq:Graph_Radio_SLAM}
           \min_{\substack{{\bf{x}}[n],\,{{\bf{u}}}_k \\ \forall n, k}} \quad \mathcal{L}.
\end{equation}
Solving this problem is still challenging since the objective function is non-linear and non-convex. To deal with this problem, we employ an iterative approach similar to the one presented in \cite{grisetti2010tutorial}, where at each iteration the problem first is locally linearized and then is solved. The algorithm then iterates until the convergence.

Let's assume that we want to optimize the following problem

\begin{equation}\label{eq:Graph_SLAM}
           \min_{\vartheta} \quad \sum_i {\bf{e}}_i^{T}(\vartheta_i) {\bf{Q}}^{-1}_i {\bf{e}}_i(\vartheta_i).
\end{equation}
where $\vartheta = [\vartheta_0^{T}, \vartheta_1^{T}, \cdots]^{T}$ is a vector of all the unknown variables, ${\bf{e}}_i(\vartheta_i)$ is a vector function of the unknown variables $\vartheta_i$, and ${\bf{Q}}_i$ is a known diagonal matrix. By using the first-order Taylor approximation around an initial guess $\breve{\vartheta_i}$, we can write
\begin{equation} \label{eq:taylor_approx}
    {\bf{e}}(\breve{\vartheta_i} + \Delta \vartheta_i) \approx \breve{{\bf{e}}}_i + {\bf{J}}_i \Delta \vartheta_i 
\end{equation}
where $\breve{{\bf{e}}}_i \triangleq {\bf{e}}(\breve{\vartheta_i})$, and ${\bf{J}}_i$ is the Jacobian of ${\bf{e}}_i(\vartheta_i)$ computed in $\breve{\vartheta_i}$. By substituting \eqref{eq:taylor_approx} in \eqref{eq:Graph_SLAM}, we  have

\begin{equation}\label{eq:Graph_SLAM_approx}
           \min_{\vartheta} \quad \sum_i \breve{{\bf{e}}}_i ^{T} \breve{{\bf{e}}}_i + 2 \breve{{\bf{e}}}_i ^{T} {\bf{Q}}^{-1}_i {\bf{J}}_i \Delta \vartheta_i + \Delta \vartheta_i^{T} {\bf{J}}_i^{T} {\bf{Q}}^{-1}_i {\bf{J}}_i \Delta \vartheta_i.
\end{equation}
We can reformulate \eqref{eq:Graph_SLAM_approx} in a matrix form as follows
\begin{equation}\label{eq:Graph_SLAM_approx_mat}
           \min_{\vartheta} \quad \breve{{\bf{e}}} + 2 {\bf{b}}^{T} \Delta \vartheta + \Delta \vartheta ^{T}\, {\bf{H}}\, \Delta \vartheta,
\end{equation}
\noindent where $\breve{{\bf{e}}} \triangleq [\breve{{\bf{e}}}_0^{T}, \breve{{\bf{e}}}_1^{T}, \cdots]^{T}$,
${\bf{b}} = [\breve{{\bf{e}}}_0 ^{T} {\bf{Q}}^{-1}_0 {\bf{J}}_0, \breve{{\bf{e}}}_1 ^{T} {\bf{Q}}^{-1}_1 {\bf{J}}_1, \cdots]^{T}$, and ${\bf{H}}$ is a block diagonal matrix defined as
\begin{equation}
   {\bf{H}} \triangleq \text{diag}\left({\bf{J}}_0^{T} {\bf{Q}}^{-1}_0 {\bf{J}}_0, {\bf{J}}_1^{T} {\bf{Q}}^{-1}_1 {\bf{J}}_1, \cdots \right).
\end{equation}
The linear problem \eqref{eq:Graph_SLAM_approx} can now be solved and the solution is given by
\begin{equation}
    \vartheta^* = \breve{\vartheta} + \Delta \vartheta^* = \breve{\vartheta} - {\bf{H}}^{-1}\, {\bf{b}},
\end{equation}
where $\breve{\vartheta}=[\breve{\vartheta}_1^{T}, \breve{\vartheta}_2^{T}, \cdots]^{T}$ is a vector of initial guesses. This procedure then repeats until $\vartheta^*$ converges to a local minima. 

With some mathematical manipulations, our original optimization problem \eqref{eq:Graph_Radio_SLAM} can be reformulated similar to \eqref{eq:Graph_SLAM}. Then it can be solved using the iterative algorithm which was mentioned above. The details of reformulating \eqref{eq:Graph_Radio_SLAM} in the form of \eqref{eq:Graph_SLAM} can be found in Appendix \ref{apx:Grapg_SLAM_Reformulation}.

\subsection{Algorithm}
The proposed algorithm iterates between two phases: 1) The EM algorithm which computes classification variables $\mathcal{W}$ and the channel parameters $\mathcal{\theta}$ by only exploiting channel gain measurements. During the first phase, the user and the UAV locations are fixed. 2) In the second phase, given the estimated parameters $\mathcal{W}, \mathcal{\theta}$, a least-squares-based SLAM framework as explained in Section \ref{sec:LS_SLAM} is used to estimate the users' locations and to track the UAV. The algorithm then iterates between these two phases until convergence (i.e. until observing
small changes in estimated value). The different steps are briefly described in Algorithm \ref{alg:Full_SLAM_EM}.

\begin{algorithm}
    \begin{algorithmic}[1]
    \STATE Given all measurements set $\mathcal{G}$.
    \STATE Initialize $\mathcal{W}, \mathcal{\theta}$, users' locations, and UAV location estimates 
    \STATE 1) \textbf{E-M algorithm}:
    \STATE  Fixing users and UAV locations estimates, find classification variables $\mathcal{W}$, and the channel parameters $\mathcal{\theta}$ using the EM algorithm.
    \STATE 2) \textbf{Least-squares-based SLAM}:
    \STATE Solve \eqref{eq:Graph_Radio_SLAM} given estimated $\mathcal{W}, \mathcal{\theta}$.
    \STATE Go to Step 3, until the change on UAV and users' locations estimate is small.
    \end{algorithmic}\caption{User localization, UAV tracking, measurement classification, and channel learning algorithm}
    \label{alg:Full_SLAM_EM}
\end{algorithm}

%% file: sections/04TrajectoryOptimization.tex
\section{UAV Trajectory Optimization}\label{sec:TrajectoryDesign}

Contrary to static anchors in wireless node localization systems, we can optimize the trajectory of the UAV (as a mobile anchor) to improve the localization performance. Specifically, UAV path planning problems that constitute a good trajectory to collect the most informative measurements, among all feasible paths, satisfying a duration or energy budget constraint can be formulated. In machine learning and robotics, such an optimization framework is often referred to as active learning or optimal design of experiments \cite{taylor2021active, PronLuc}. The relevance of this problem to our localization scenario can be understood as follows: The measurements collected from NLoS links usually result in a degradation of the localization accuracy due to the higher shadowing effect and the bias term in timing measurements for NLoS channels. On the other hand, devising a trajectory for the UAV to maintain LoS links to all users at all times is not a viable solution since there may not exist a continuous trajectory that fulfills this constraint and because of the limited mission time. Therefore, optimizing the UAV trajectory to strike a balance between collecting LoS measurements from the users and the mission time is crucial.

To do so, in this section, we employ the notion of Fisher information matrix (FIM) \cite{amari2007methods} to find a maximally informative UAV trajectory. This metric helps in measuring the amount of information that collected measurements carry about unknown user locations.
Our goal is to exploit the structural properties of the FIM, so as to design an optimal policy for the drone to collect the best possible measurements from users. In the following, we briefly explain the FIM and its properties and then optimize the UAV trajectory.

Note that the UAV trajectory is optimized to collect most informative measurements from users, hence in this section we do not consider the measurements collected by the BSs. Moreover, we only focus on the ToA measurements, since the radio link is the same at each time for collected ToA and channel gain measurements.

\subsection{Fisher Information Matrix}
For a set of ToA measurements collected by the UAV from users, the FIM of the measurements with respect to the users' locations is given by
 \begin{equation}
     {\bf{F}} = \Exp\left[ \frac{\partial\mathcal{L}_{\text{UAV-UE}}^{\tau} }{\partial {\bf{u}}} {\frac{\partial\mathcal{L}_{\text{UAV-UE}}^{\tau}}{\partial {\bf{u}}}}^\Tr \right],
 \end{equation}
where ${\bf{u}}=[{\bf{u}}_1^{T},\cdots,{\bf{u}}_K^{T}]^{\Tr}$ is a vector stacking all the users locations, and $\mathcal{L}_{\text{UAV-UE}}^{\tau}$ is the log-likelihood of ToA measurements collected by the UAV from users defined as
 \begin{equation}
 \begin{aligned}
     \mathcal{L}_{\text{UAV-UE}}^{\tau, s} \triangleq & \sum_{n=1}^{N}\sum_{k=1} ^ K\log\, w_k[n] f_{\tau, k,\text{LoS}}[n] \, + \\
     & \sum_{n=1}^{N}\sum_{k=1} ^ K\log\, (1-w_k[n]) f_{\tau, k,\text{NLoS}}[n],
      \end{aligned}
 \end{equation}
where $f_{\tau, k,s}[n]$ is defined in \eqref{eq:PDF_ToA}. Then the FIM for all measurements collected up to time step $N$ is given by 
 \begin{equation}
 \begin{aligned}
          {\bf{F}}_{N} =& \sum_{n=1}^{N}\sum_{k=1}^K {\bf{H}}_{k}[n]\\
          =& \,{\bf{F}}_{N-1} + \sum_{k=1}^K {\bf{H}}_{k}[N], \label{eq:FIM_equation}
 \end{aligned}
 \end{equation}
 where 
 \begin{equation} \label{eq:H_matrix_segmented}
     {\bf{H}}_{k}[n] = w_k[n] {\bf{H}}_{k,\text{LoS}}[n] + (1-w_k[n]) {\bf{H}}_{k,\text{NLoS}}[n], 
 \end{equation}
and ${\bf{H}}_{k,s}[n]$ can be derived akin to the work in \cite{esrafilian20203d}. Note that, \eqref{eq:FIM_equation} implies that the FIM is cumulative over time.
 \subsection{Cramér-Rao Bound Analysis}
According to  the Cramér-Rao bound (CRB) \cite{van2007parameter}, the mean squared error (MSE) of the estimated parameters ${\bf{u}}$ given the measurements for an unbiased estimator is lower bounded by 
   \begin{equation}
 \begin{aligned}
          \text{MSE}({\bf{u}})\ge\trace{({\bf{F}}_{N}^{-1})}, \label{eq:MSE_CRB_LB}
 \end{aligned}
 \end{equation}

\noindent where $\trace{({\bf{F}}_{N}^{-1})}$ is the trace of ${\bf{F}}_{N}^{-1}$. Using the matrix inversion lemma, it can be shown that ${\bf{F}}_{N,s}^{-1}$ follows a recursive relation as follows
 \begin{equation} \label{eq:FIM_recursion}
     {\bf{F}}_{N}^{-1} = {\bf{F}}_{1}^{-1} - \sum_{n=2}^{N}{\bf{R}}[n],
 \end{equation}
 where ${\bf{R}}[n]$ is defined as the amount of improvement in the estimation within time slot $n$ which is given by
 \begin{equation} \label{eq:recursive_award_function}
     {\bf{R}}[n] = {\bf{F}}_{n-1}^{-1}\left(\sum_{k=1}^K{\bf{H}}_{k}[n]^{-1}+{\bf{F}}_{n-1}^{-1}\right)^{-1}{\bf{F}}_{n-1}^{-1}.
 \end{equation}
 This recursion will be used later to optimize the UAV trajectory.
\subsection{Trajectory Optimization}
We are seeking to devise a trajectory for the UAV under a limited flight mission time during which the UAV starts from ${\bf{x}}_{\text{I}}$ and ends up at the terminal point ${\bf{x}}_{\text{F}}$ while minimizes the estimation error of parameters ${\bf{u}}$. Such optimization problem can be formulated as
\begin{subequations}\label{eq:TRJ_OPT_Origin}
 \begin{align}
  \begin{split}
          \min_{\{{\bf{x}}[n]\}_{n=1}^N} &\quad \text{MSE}({\bf{u}}) \label{eq:TRJ_OPT_Origin_O.F.}
 \end{split}\\
   \begin{split}
          \text{s.t.}&\quad
           {\bf{x}}[1] = {\bf{x}}_{\text{I}},\, {\bf{x}}[N] = {\bf{x}}_{\text{F}}, \label{eq:TRJ_OPT_Origin_x_t}
 \end{split}\\
 \begin{split} \label{eq:TRJ_OPT_max_speed}
           &\quad \| {\bf{x}}[n] - {\bf{x}}[n-1] \| \le D_{\text{max}}, \, n \in [2, N],
 \end{split}
 \end{align}
\end{subequations}
where $D_{\text{max}}$ is the maximum distance that the UAV can travel within each time step. The optimization problem \eqref{eq:TRJ_OPT_Origin} is challenging to solve since a closed form expression for $\text{MSE}({\bf{u}})$ is not known. Therefore, we instead find a trajectory to minimize the CRB. Note that despite the fact that CRB in \eqref{eq:MSE_CRB_LB} only applies for unbiased estimators, in this case, it nevertheless provides a good metric for finding favorable measurements in our trajectory optimization.
Using the CRB in \eqref{eq:MSE_CRB_LB}, we can approximate \eqref{eq:TRJ_OPT_Origin}
as

   \begin{subequations}\label{eq:TRJ_OPT_LB_Rec}
 \begin{align}
  \begin{split}
          \min_{\{{\bf{v}}[n]\}_{n=1}^N} &\quad \trace{({\bf{F}}_{N}^{-1})} 
 \end{split}\\
    \begin{split}
          \text{s.t.}&\quad \eqref{eq:TRJ_OPT_Origin_x_t}, \eqref{eq:TRJ_OPT_max_speed}.
 \end{split}
  \end{align}
  \end{subequations}
Even after this approximation of the original problem, \eqref{eq:TRJ_OPT_LB_Rec} is still difficult to solve. The matrix 
${\bf{F}}_{N}^{-1}$ is a function of the unknown parameters ${\bf{u}}$ (that need to be estimated), hence
it can not be computed exactly. To cope with this problem, we take a sequential approach where at time step $n$, we use an estimate of ${\bf{F}}_{N}^{-1}$, denoted by $\hat{\bf{F}}_{N|n}^{-1}$, as the objective to be minimized. This estimate is obtained from using 
$\hat{{\bf{u}}}_n$, which represents an estimate of ${\bf{u}}$
obtained from measurements collected up to the time slot $n$.
So essentially, \eqref{eq:TRJ_OPT_LB_Rec} becomes an online learning and trajectory design problem where at each time step during the mission after obtaining new measurements, unknown parameter estimates are updated, and accordingly, a new trajectory from that point is generated. In the following, we elaborate on an online and low-complexity greedy algorithm to find a sub-optimal trajectory.

  \subsection{Greedy Trajectory Design} \label{sec:greedy_trj_design}
In this section, we propose a greedy approach where the trajectory is devised locally and piece by piece. At each time step, the UAV aims to find the best location for the next step to go and collect measurements that can potentially offer the maximum improvement in the estimation error. Let's define the objective function in the greedy algorithm at time step $n$ as
 \begin{equation} \small
      L({\bf{x}}[n]) = \left\{\begin{array}{rcl}
        \trace{(\hat{\bf{R}}_{n|n-1})}, & \|{\bf{x}}[n]-{\bf{x}}_{\text{F}}\|\le \lambda\\
        -\infty ,& \text{otherwise}
      \end{array}\right. ,\label{eq:Greedy_Cost}
  \end{equation}
where $\hat{\bf{R}}_{n|n-1}$ is the estimation of ${\bf{R}}[n]$ given $\hat{{\bf{u}}}$ up to step $n$, and $\lambda=D_{\text{max}} (N-n)$.
Note that the relation between $\hat{\bf{R}}_{n|n-1}$ and $\hat{\bf{F}}_{N|n-1}$ can be inferred from \eqref{eq:FIM_recursion}.  
The definition in \eqref{eq:Greedy_Cost} forces the UAV to reach the terminal point ${\bf{x}}_{\text{F}}$ within the total flying time constraint.

Finally, the next optimal drone position at any time step $n\in [1,N-1]$ can be obtained by solving
\begin{equation} \label{eq:Re_Greedy_TRJ_UAV_model}
\begin{aligned}
\max_{{\bf{x}}[n+1]}
&\quad L({\bf{x}}[n+1])\\
\text{s.t.} 
& \quad {\bf{x}}[1] = {\bf{x}}_{\text{I}},\\
& \quad \| {\bf{x}}[n+1] - {\bf{x}}[n] \| \le D_{\text{max}}.
\end{aligned} 
\end{equation}

\begin{figure}[t]
\begin{centering}
\includegraphics[width=0.4\columnwidth]{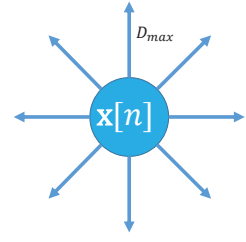}
\par\end{centering}
\caption{An example of the possible actions in the greedy trajectory design at the $n$-th time step.  
\label{fig:Greedy_TRJ_design}}
\end{figure}

\noindent To solve problem \eqref{eq:Re_Greedy_TRJ_UAV_model}, we initialize $n=1$ and the drone starts flying from base point ${\bf{x}}[1] = {\bf{x}}_{\text{I}}$. To find the best drone position in the next time step $({\bf{x}}[n+1], n\in [1,N-1])$, we discretize the search space around the current drone location and we calculate \eqref{eq:Greedy_Cost} for all adjacent points. Then the adjacent locations with the maximum value is chosen as the drone location in the next step. In accordance with \eqref{eq:H_matrix_segmented} and \eqref{eq:recursive_award_function}, the objective $L({\bf{x}}[n+1])$ is a function of LoS/NLoS labels of measurements which will be collected in the future time step $n+1$. To estimate the labels of the future measurements, we use the LoS probability introduced in \cite{AlhKaLa} which is given by

\begin{equation}
    w_{k}[n+1]=\frac{1}{1+\exp\left(a\,\psi_{k}[n+1]+b\right)},
\end{equation}

\noindent where $\psi_{k}[n+1]=\arctan(z[n+1]/r_{k}[n+1])$ denotes the elevation angle and $r_{k}[n+1]$ is the ground projected distance between the UAV at time step $n+1$ and the $k$-th user. Parameters $\left\{ a,b\right\} $ are the model coefficients which are computed according to \cite{AlhKaLa} and based on the characteristics of the city.

Note that if the value of all the neighbor locations are computed as infinity, then the UAV moves towards the terminal location ${\bf{x}}_{\text{F}}$ by $\frac{\|{\bf{x}}[n]-{\bf{x}}_{\text{F}}\|}{N-n}$ meters, where $n$ is the current time step. 
In Fig. \ref{fig:Greedy_TRJ_design}, an example of the greedy trajectory design at the $n$-th time step is shown. To find the best position for the UAV, at the next time step, all eight UAV's neighbor positions at time step $n$ need to be evaluated.

%% file: sections/05results.tex
\begin{figure}[t]
\centering
\subfloat[]{
  \includegraphics[clip,width=0.9\columnwidth]{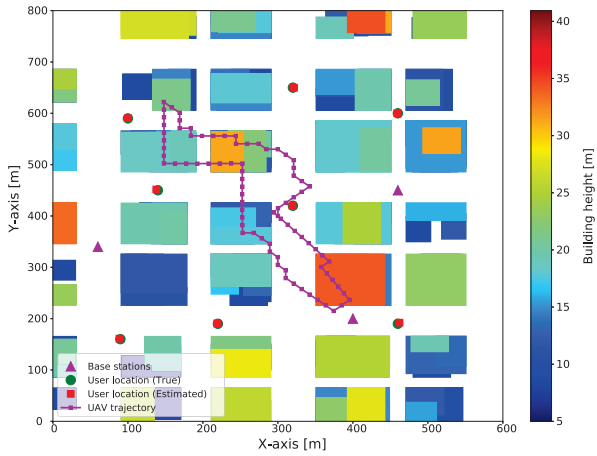}
}
\newline
\centering
\subfloat[]{
  \includegraphics[clip,width=0.9\columnwidth]{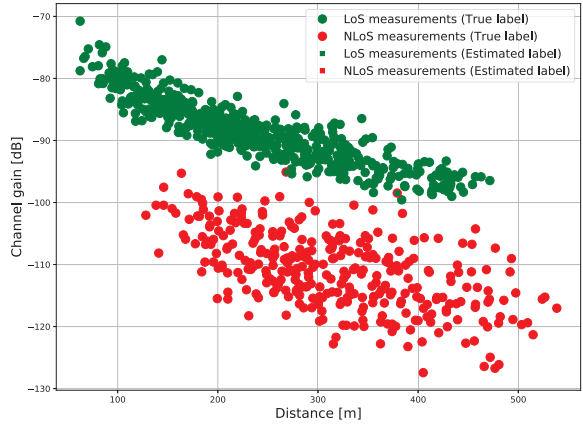}
}
\caption{(a) The performance of proposed localization algorithm for multi-user case and the optimized UAV trajectory, (b) Corresponding channel gain measurements collected from users and the result of the classification. The classification error is 0.2 $\%$. \label{fig:top_view}}
\end{figure}

\begin{figure}[t]
\begin{centering}
\includegraphics[width=0.9\columnwidth]{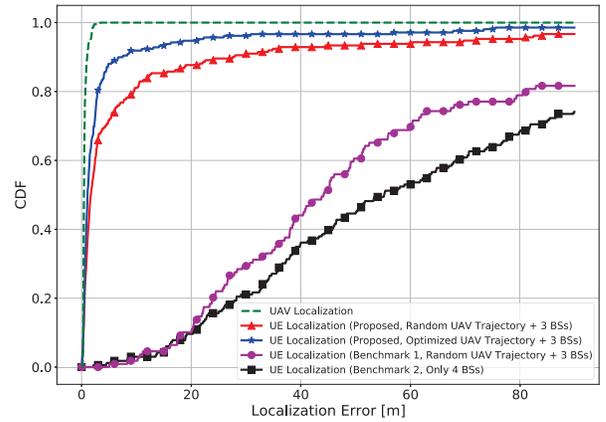}
\par\end{centering}
\caption{The CDF of user localization error for different algorithms.  
\label{fig:cdf_localization}}
\vspace{-6pt}
\end{figure}

\begin{figure}[t]
\begin{centering}
\includegraphics[width=0.8\columnwidth]{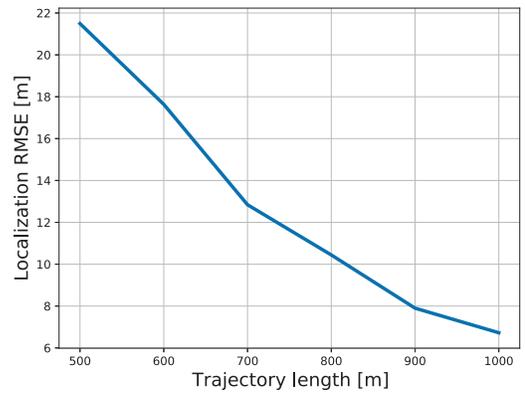}
\par\end{centering}
\caption{The proposed localization error in terms of RMSE versus increasing the UAV trajectory length.  
\label{fig:RMSE_trj_len}}
\vspace{-6pt}
\end{figure}

\section{Numerical Results}\label{sec:simulations}
A dense urban city neighborhood comprising buildings and streets as shown in Fig. \ref{fig:top_view}-a is considered. The height of the buildings is Rayleigh distributed in the range of 5 to \SI{40}{m} \cite{Ref26_HourKandeepJamail}. The true propagation parameters are chosen as  $\alpha_{\text{LoS}}=-22,\,\alpha_{\text{NLoS}}=-32,\,\ss_{\text{LoS}}=-32\,\text{dB},\,\ss_{\text{NLoS}}=-35\,\text{dB}$
according to an urban micro scenario \cite{3GPP}. The variances
of the shadowing components in \text{LoS} and \text{NLoS} scenarios are
$\sigma_{\text{LoS}}^{2}=2\,\text{dB}$, and $\sigma_{\text{NLoS}}^{2}=5\,\text{dB}$, respectively. For the ToA measurements, the true parameters are chosen as $\mu_{\tau, \text{LoS}} = 0, \, \sigma^2_{\tau, \text{LoS}} = 2 \text{ m}, \mu_{\tau, \text{NLoS}} = 50 \text{ m}, \, \sigma^2_{\tau, \text{NLoS}} = 40 \text{ m}$, which is roughly equivalent to $100$ MHz bandwidth. We assume that there are three fixed BSs in the city, as shown in Fig. \ref{fig:top_view}-a , with the same altitudes of $25 \text{ m}$. The altitude of the UAV is assumed to be fixed and set to $80 \text{ m}$, and ${\bf{x}}_{\text{I}} = {\bf{x}}_{\text{F}}= [300, 400, 80]^T$. The GPS and the IMU each has a covariance of $\sigma_{gps}^2 = 5 \text{ m}, \sigma_{vel}^2 = 0.2 \text{ m/s}$.

In Fig. \ref{fig:top_view}-a, the results of proposed algorithm for multi-user case ($K=8$) is shown. To localize the users, the UAV trajectory is optimized with the maximum trajectory length of $1000 \text{ m}$. The UAV tries to visit the users which experience the worst conditions (more likely to be NLoS) in order to improve the localization performance, and it can be seen that the users are accurately localized (the average localization accuracy is $1.5 \text{ m}$). The results of the measurement classifications are shown in Fig. \ref{fig:top_view}-b. For ease of exposition, we only showed the classified channel gain measurements. We can see that the algorithm could classify the measurements with a low error of 0.2 $\%$.

In Fig. \ref{fig:cdf_localization}, the cumulative distribution function (CDF) of user localization error of our proposed algorithm with comparison to different benchmarks over Monte-Carlo simulations is shown. The length of the UAV trajectory is set to $800 \text{ m}$. In benchmark 1, only RSS measurements are used to localize the users and track the UAV, this benchmark can be considered as an extension of works in \cite{esrafilian20203d, YinFritGus}. In benchmark 2, only 4 static BSs are used (no UAV). A rectangular trajectory for the UAV with a similar length of $800 \text{ m}$ is chosen as the random trajectory (as opposed to the optimized trajectory). It is also clear that optimizing the UAV trajectory can considerably improve the localization performance. From benchmark 2, we can see the gain brought by using mobile anchors in localization systems performance. We have also shown the results of UAV tracking (green dashed line), we can see that the UAV can be tracked more accurately since it can always maintain LoS connections to the BSs.

In Fig. \ref{fig:RMSE_trj_len},  the performance of the proposed algorithm, in terms of root-mean-square error (RMSE), is shown when the length of the UAV trajectory increased over several Monte-Carlo simulations. The accuracy of user localization increases when the length of the UAV trajectory increases. This stems from the fact that by increasing the UAV trajectory the chance of collecting more LoS measurements increases which results in better localization performance.

%% file: sections/06Appendix.tex
\appendix
\subsection{Reformulation of Problem \eqref{eq:Graph_Radio_SLAM} to a Vector Form} \label{apx:Grapg_SLAM_Reformulation}
To reformulate problem \eqref{eq:Graph_Radio_SLAM} into a vector representation, we first define $\vartheta$ as follows

\begin{equation}
    \vartheta =\left [{\bf{x}}[1]^{T}, \cdots, {\bf{x}}[N]^{T}, {\bf{u}}_1^{T}, \cdots^{T}, {\bf{u}}_K^{T} \right]^T.
\end{equation}

having defined $\vartheta$, we can define $\mathcal{L}_{\text{UAV}}$ as follows

\begin{equation}
    \mathcal{L}_{\text{UAV}} = {\bf{e}}_{gps}^T  {\bf{Q}}_{gps}^{-1} {\bf{e}}_{gps} + {\bf{e}}_{vel}^T  {\bf{Q}}_{vel}^{-1} {\bf{e}}_{vel},
\end{equation}
where ${\bf{e}}_{gps} \triangleq \left [\hat{{\bf{x}}}^{T}[1] - {\bf{x}}^{T}[1], \cdots, \hat{{\bf{x}}}^{T}[N] - {\bf{x}}^{T}[N]\right]^T$, ${\bf{e}}_{vel} \triangleq \left [\hat{{\bf{v}}}^{T}[2] - {\bf{v}}^{T}[2], \cdots, \hat{{\bf{v}}}^{T}[N] - {\bf{v}}^{T}[N]\right]^T$, ${\bf{Q}}_{gps} \triangleq \sigma_{gps}^2 * {\bf{I}}_N$, and ${\bf{Q}}_{vel} \triangleq \sigma_{vel}^2 * {\bf{I}}_{N-1}$, where ${\bf{I}}_{n}$ is the identity matrix of size $n \times n$.

To reformulate $\mathcal{L}_{\text{UAV-UE}}$ we ignore the terms which do not depend on UAV or users' locations. Hence we can write

\begin{equation}
 \begin{aligned}
          \mathcal{L}_{\text{UAV-UE}} &= {\bf{e}}_{g, \textbf{LoS}}^T \, {\bf{Q}}_{g, \textbf{LoS}} ^{-1} \, {\bf{e}}_{g, \textbf{LoS}} + \\
         & {\bf{e}}_{g, \textbf{NLoS}}^T \, {\bf{Q}}_{g, \textbf{NLoS}} ^{-1} \, {\bf{e}}_{g, \textbf{NLoS}} +\\
          & {\bf{e}}_{\tau}^T \, {\bf{Q}}_{\tau, \textbf{LoS}}^{-1} \, {\bf{e}}_{\tau} +\\
          & {\bf{e}}_{\tau}^T \, {\bf{Q}}_{\tau, \textbf{NLoS}}^{-1} \, {\bf{e}}_{\tau},
      \end{aligned}
 \end{equation}
 
\noindent where ${\bf{e}}_{g, s} \triangleq \left [\hat{g}_{1}[1] - g_{1, s}[1], \cdots, \hat{g}_{K}[N] - g_{K, s}[N] \right]^T \quad$, ${\bf{e}}_{\tau} \triangleq \left [\hat{\tau}_{1}[1] - \tau_{1}[1], \cdots, \hat{\tau}_{K}[N] - \tau_{K}[N] \right]^T$, and
\begin{equation}
 \begin{aligned}
       {\bf{Q}}_{g, \text{LoS}}^{-1} &\triangleq \text{diag} \left( \frac{w_1[1]}{\sigma_{\text{LoS}}^2}, \cdots,  \frac{w_K[N]}{\sigma_{\text{LoS}}^2} \right),\\
        {\bf{Q}}_{g, \text{NLoS}}^{-1} &\triangleq \text{diag} \left( \frac{1-w_1[1]}{\sigma_{\text{NLoS}}^2}, \cdots,  \frac{1-w_K[N]}{\sigma_{\text{NLoS}}^2} \right),\\
        {\bf{Q}}_{\tau, \text{LoS}}^{-1} &\triangleq \text{diag} \left( \frac{w_1[1]}{\sigma_{\tau, \text{LoS}}^2}, \cdots,  \frac{w_K[N]}{\sigma_{\tau, \text{LoS}}^2} \right),\\
        {\bf{Q}}_{\tau, \text{NLoS}}^{-1} &\triangleq \text{diag} \left( \frac{1-w_1[1]}{\sigma_{\tau, \text{NLoS}}^2}, \cdots,  \frac{1-w_K[N]}{\sigma_{\tau, \text{NLoS}}^2} \right).
      \end{aligned}
 \end{equation}
 
$\mathcal{L}_{\text{BS-UAV}}$, and $\mathcal{L}_{\text{BS-UE}}$ can also be reformulated in a similar manner as above.